\documentclass[sigconf]{acmart}
\usepackage{tabu}
\usepackage{booktabs}
\usepackage{graphicx}

\graphicspath{{./}}
\renewcommand\footnotetextcopyrightpermission[1]{}

\begin{document}

\title{GBake: Baking 3D Gaussian Splats into Reflection Probes}

\author{Stephen Pasch}
\affiliation{%
  \institution{Columbia University}
  \city{New York}
  \state{NY}
  \country{USA}
}
\email{ssp2188@columbia.edu}

\author{Joel K. Salzman}
\affiliation{%
  \institution{Columbia University}
  \city{New York}
  \state{NY}
  \country{USA}
}
\affiliation{%
  \institution{Brown University}
  \city{Providence}
  \state{RI}
  \country{USA}
}
\email{joel.salzman@columbia.edu}

\author{Changxi Zheng}
\affiliation{%
  \institution{Columbia University}
  \city{New York}
  \state{NY}
  \country{USA}
}
\email{cz2280@columbia.edu }

\copyrightyear{2025}
\acmYear{2025}
\acmConference{SIGGRAPH Posters '25}{August 10-14, 2025}{Vancouver, BC, Canada}\acmBooktitle{Special Interest Group on Computer Graphics and Interactive Techniques Conference Posters (SIGGRAPH Posters '25), August 10-14, 2025}\acmDOI{10.1145/3721250.3742978}
\acmISBN{979-8-4007-1549-5/2025/08}

\begin{abstract}
 The growing popularity of 3D Gaussian Splatting has created the need to integrate traditional computer graphics techniques and assets in splatted environments. Since 3D Gaussian primitives encode lighting and geometry jointly as appearance, meshes are relit improperly when inserted directly in a mixture of 3D Gaussians and thus appear noticeably out of place. We introduce GBake, a specialized tool for baking reflection probes from Gaussian-splatted scenes that enables realistic reflection mapping of traditional 3D meshes in the Unity game engine.

\end{abstract}



\begin{teaserfigure}
  \centering
  \includegraphics[width=0.98\linewidth]{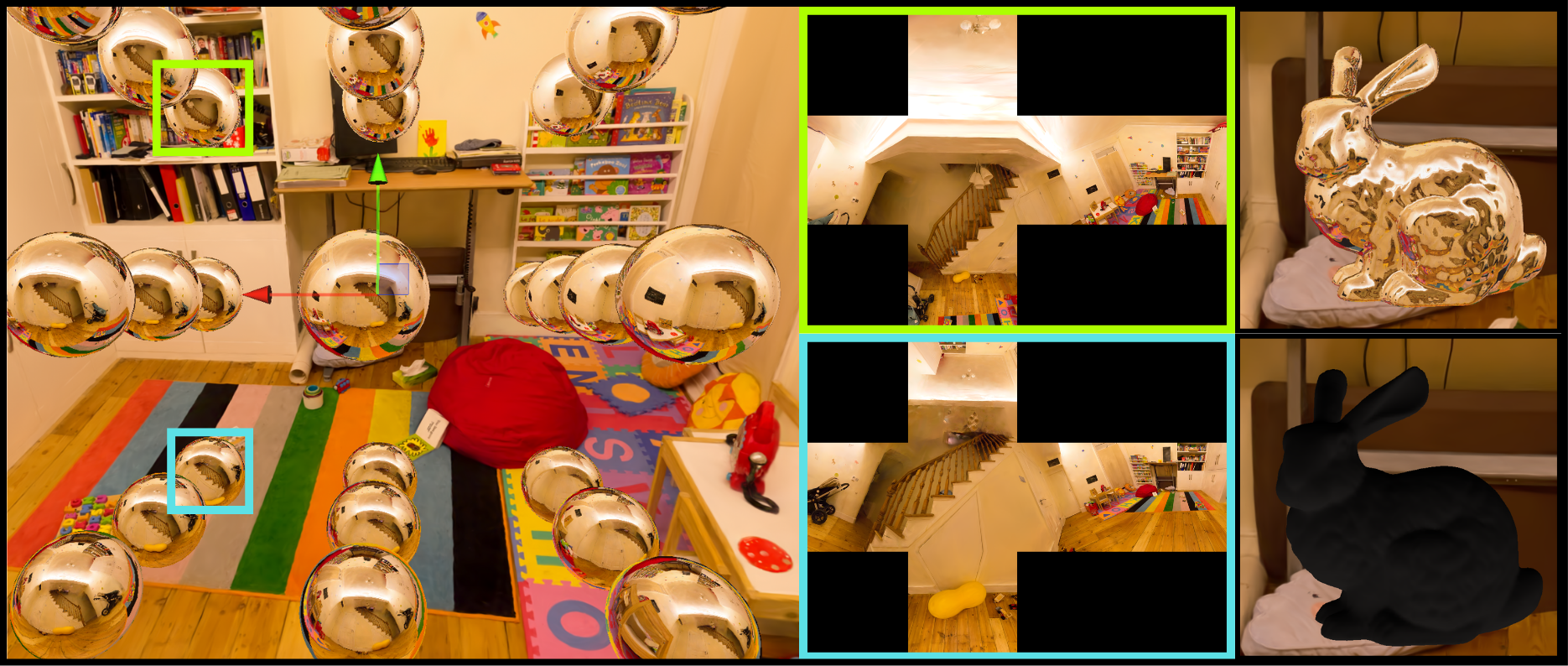}
  \caption{GBake enables mesh reflections in a gaussian-splatted environment. \textbf{Left:} Visualization of GBake reflection probes. 
  \textbf{Middle:} Corresponding cubemaps of the highlighted probes. \textbf{Upper Far Right:} Highly specular metallic mesh object utilizing the baked probes. \textbf{Lower Far Right:}  The same object with the probes disabled, demonstrating absence of environmental detail.}
  \Description{A mesh object reflects its surrounding Gaussian splat environment using cubemaps generated with GBake. The object is shown with and without the baked probes enabled.}
  \Description{A mesh object reflects its surrounding Gaussian splat environment using cubemaps generated with GBake. The object is shown with and without the baked probes enabled.}
  \label{fig:teaser}
\end{teaserfigure}

\maketitle

\begin{center}
    \fbox{\parbox{0.97\linewidth}{
    \small
    \textcopyright~Stephen Pasch, Joel Salzman, Changxi Zheng | ACM 2025. 
    This is the author's version of the work. It is posted here for your personal use. Not for redistribution.\\
    The definitive Version of Record was published in \textit{Special Interest Group on Computer Graphics and Interactive Techniques Conference Posters (SIGGRAPH Posters '25), August 10-14, 2025},\\
    \url{http://dx.doi.org/10.1145/3721250.3742978}
    }}
\end{center}

\section{Introduction}
\begin{figure*}[!t]
    \centering
    \includegraphics[width=0.75\textwidth]{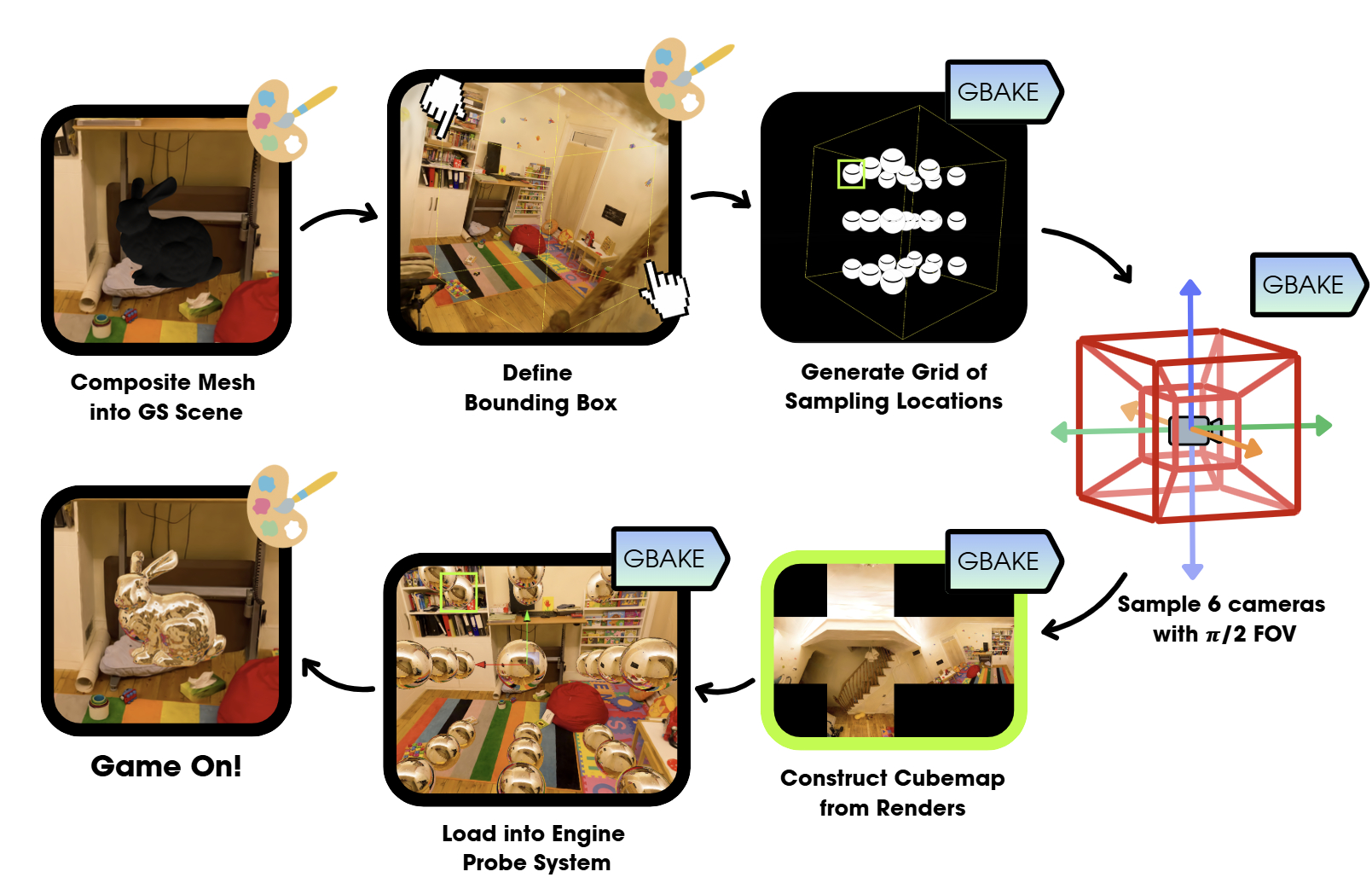}
    \caption{ Gbake pipeline. Each stage is distinguised by what is in the artists control(paint swatch) and what the system handles(Gbake). }
    \label{fig:pipeline}
    \Description{The pipeline of GBake. In order it is Composite Mesh into GS Scene, define bounding box, generate grid of sampling locations, sample 6 camera with pi/2, construct cubemap from renders, Load into Engine Probe System, Game on!}
\end{figure*}
Since the advent of 3D Gaussian Splatting (hereafter \textit{3DGS}) by Kerbl et al.~\cite{kerbl3Dgaussians}, there has been increasing desire to use traditional assets inside Gaussian-based 3D reconstructions. The difficulty with using 3D Gaussians in traditional graphics pipelines is that volumetric particle rasterization is a fundamentally different paradigm from rendering techniques that model light transport by tracing rays over multiple bounces. Since 3DGS optimizes appearance, the resulting scenes contain neither explicit light emitters nor surface geometry. This inhibits common graphics techniques such as relighting and reflection mapping for meshes introduced to a splatted scene. 

We address this limitation with GBake, a tool that bakes 3DGS scenes into a set of environment maps (alternatively called reflection probes). These probes enable realistic reflection mapping of arbitrary mesh-based objects inserted into the scene, effectively bridging the gap between appearance-optimized Gaussian representations and traditional mesh rendering pipelines. We demonstrate GBake's utility across multiple trained 3D Gaussian scenes by introducing and relighting meshes in the Unity game engine \cite{Unity}.

Properly baking a scene requires multiple environment maps to encode the heterogeneity of lighting within the scene. Otherwise, an object surface will be lit by the same set of incoming rays irrespective of its location. This causes obvious errors when light sources are near the relit object. For this reason, GBake creates a regular grid of reflection probes inside the scene, enabling its use in a diverse array of graphics applications.

\section{Related Work}
\textbf{\textit{Precomputed Lighting}} Precomputed lighting techniques are fundamental to real-time rendering, with lightmaps~\cite{RamamoorthiLightmap01}, spherical harmonics~\cite{Kautz02,SloanStupidSH}, and precomputed radiance transfer~\cite{SloanPRT02} being widely adopted in interactive applications. Our work specifically builds on environment mapping techniques~\cite{DebevecReflectanceMapping}, where cubemaps capture environmental lighting at discrete locations. Incident light at arbitrary points can then be estimated by interpolating nearby environment maps. This enables realistic reflections on meshes without explicitly simulating light transport for each fragment.

\textbf{\textit{Training 3D Gaussians}}. Scenes comprised of 3D Gaussians are usually trained via splatting \cite{kerbl3Dgaussians}. In our comparative analysis, we use Splatfacto~\cite{xu2024splatfactownerfstudioimplementationgaussian}, an open-source implementation of 3DGS that also employs EWA Splatting~\cite{1021576} to render trained scenes. 3DGS approximates the 2D footprint of the 3D particle with respect to the image plane as a function of the Gaussian's rotation and scale as \begin{math}\Sigma = RSS^{T}R^{T}\end{math}. These values are alpha-blended to contribute colors to the pixels intersecting the footprint.

GBake is built on 3D Gaussian Raytracing (hereafter \textit{3DGRT})~\cite{3dgrt2024}, which employs raytracing rather than splatting to render 3D Gaussian representations. 3DGRT leverages the DirectX raytracing library to calculate pixel colors via volume rendering. The contribution of an intersecting particle to a ray is calculated as the maximum opacity on the ray-particle intersection. This function is independent of the camera orientation and instead depends on the ray direction and particle attributes. Although scenes can also be trained with 3DGRT, we only bake splatted scenes since they are more common.

\begin{figure}[t]
    \centering
    \includegraphics[width=0.40\textwidth]{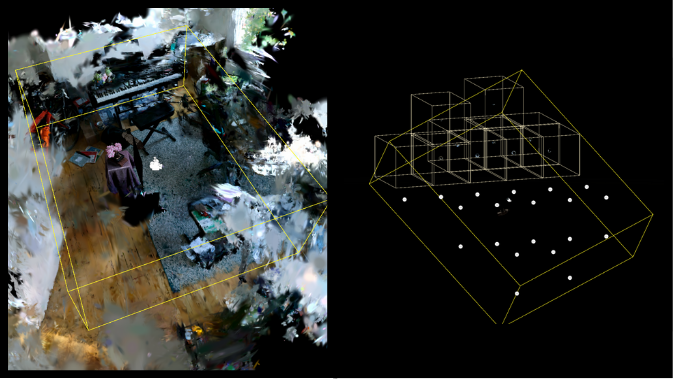}
    \caption{ \textbf{Left}: Artist defines bounding box in editor (yellow). \textbf{Right}: Baked reflection probes with influence zones (tan).}
    \label{fig:Editor}
    \Description{The left image is a gaussian splatted scene with a yellow bounding box within it. The right image is that same bounding box with gizmos visualizing the location of sampled reflection probes from GBake.}
\end{figure}

\section{Implementation}

Our baking system generates a 3D grid of probe locations within a bounding volume. At each probe location, we render six perspective views sharing the same origin, facing along the primary axes ($+X$, $+Y$, $+Z$, $-X$, $-Y$, $-Z$) with a viewing angle of ${\pi}/{2}$. These six views are composited into cubemaps and exported as PNG files. A user can manually define a bounding box and specify the grid resolution along each axis. Once baking is complete, the system generates a JSON file containing the cubemap filepaths and the reflection probes' spatial locations, which are then loaded into Unity as reflection probes. Each probe's influence volume is calculated based on inter-probe distances with an artist-controlled overlap between adjacent probes to ensure smooth transitions as objects move through the scene (Fig.~\ref{fig:Editor}). This approach enables each point on a mesh surface to sample and interpolate from nearby reflection probes, facilitating physically plausible reflections as an object moves from one zone to the next. Our implementation consists of Python code for the baking and a Unity interface for visualization and artist control~\cite{UnityGaussianSplatting}.

\begin{figure}[b]
    \includegraphics[width=0.40\textwidth]{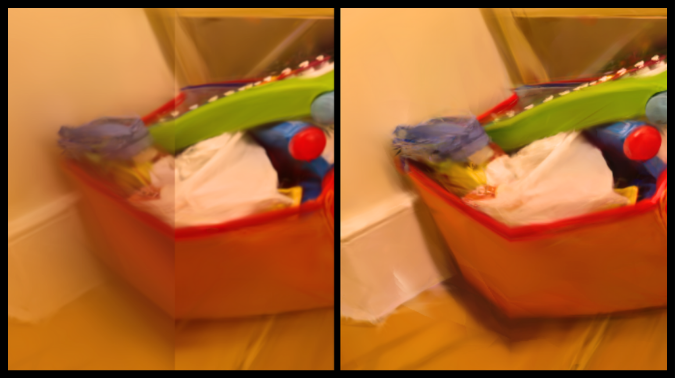}
    \caption{Comparison of cubmap seams between faces. \textbf{Left:} Splatted face intersections show color discontinutity. \textbf{Right:} Raytraced cubemap faces eliminate visible discontinuities.}
    \Description{Two cubemaps are shown: one with visible seams between faces generated through splatting and one with smooth transitions generated through raytracing.}
    \label{fig:splatvtrace}
\end{figure}
\section{Raytracing versus Splatting}
Our original intention was to bake scenes by splatting. We conducted extensive experiments with this technique using Splatfacto. However, the output consistently demonstrated artifacting at the seams of the cubemaps (Fig.~\ref{fig:splatvtrace}). Specifically, we found misalignment in the footprints of the same Gaussian particles when splatted to two different cameras. This resulted in obvious color discontinuities at the intersection of the relevant image planes (cube edges), which are unacceptable in a cubemap because the orientation of the cube should be independent of the incident light.

We conjecture that this results from the approximation used in EWA Splatting \cite{1021576}, the rasterization algorithm upon which 3DGS and Splatfacto rely. EWA Splatting is purposefully approximate for the sake of speed. Its error term is discussed in detail by Huang et. al. \cite{huang2024erroranalysis3dgaussian}. Let $\theta_\mu, \phi_\mu$ be the 3D polar coordinates of the angle made between a particle mean $\mu$ and the pose vector of the camera to which it is splatted. We highlight that the full closed-form error expression $\epsilon(\theta_\mu, \phi_\mu)$ contains irreducible terms of $sin(\theta_\mu)$, $cos(\theta_\mu)$, $tan(\theta_\mu)$, $sin(\phi_\mu)$, $cos(\phi_\mu)$, and $tan(\phi_\mu)$. By construction, a particle $\mu$ on the seam of a cubemap is simultaneously splatted with respect to a distinct optical axis for each incident cube face. Clearly, the angles $\theta_\mu$ and $\phi_\mu$ differ depending on the camera, so the errors $\epsilon$ differ, and thus the calculated colors will be inconsistent.

In contrast, raytracing utilizes no approximation relative to the orientation of the camera. Two rays with identical origins and identical directions will contribute identical colors. This is apparent in the smooth color transition between cubemap faces as shown in Fig.~\ref{fig:splatvtrace}. The tradeoff is that raytracing is significantly slower; for baking, we prioritize quality over speed.
\begin{figure}[t]
    \centering
    \includegraphics[width=0.45\textwidth]{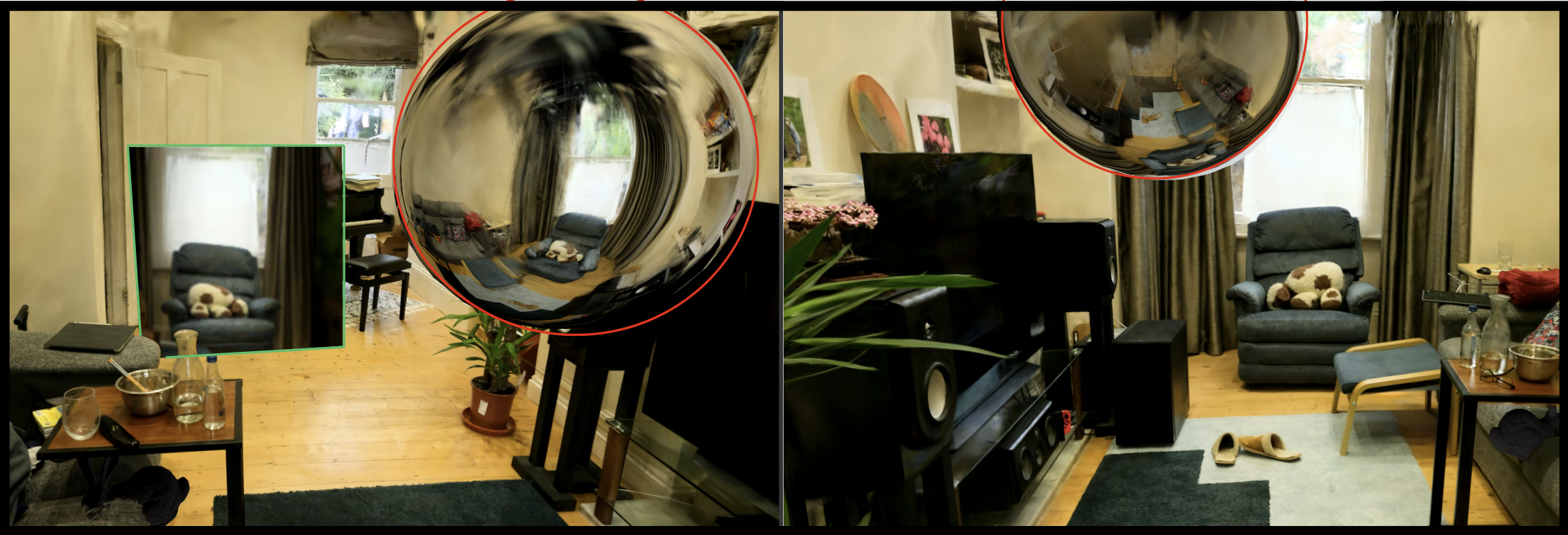}
    \caption{Two vantage points in a baked scene. \textbf{Left}: A planar (outlined in green) and spherical (outlined in red) mirror added into the scene. \textbf{Right}: The same spherical mirror from another vantage point.}
    \label{fig:room}
    \Description{A 3DGS scene with two chrome surfaces injected into it. One is a square plane and the other is a sphere. }
\end{figure}

\begin{figure}[h]
  \centering
    \renewcommand{\thetable}{\arabic{table}}
    \renewcommand{\arraystretch}{1.31}
    \small
    \captionof{table}{GBake scene baking performance. Bounding boxes in interior regions were manually inputted for each scene.}
    \begin{tabular}{lcc}
      \toprule
      \textbf{Dataset} & \textbf{Gaussians (\#)} & \textbf{Runtime (s)} \\
      \midrule
      Bonsai & 1,244,819 & 21.08 \\
      \midrule
      DrJohnson & 3,177,554 & 27.27 \\
      \midrule
      Playroom & 1,916,379 & 35.86 \\
      \midrule
      Room & 1,593,376 & 26.27 \\
      \bottomrule
    \label{tab:scene_comparison}
    \vspace{-15pt}
    \end{tabular}
    \vspace{-15pt}
\end{figure}

\section{Experiments}
We evaluate GBake across multiple pretrained scenes from the Mip-Nerf 360 dataset ~\cite{barron2022mipnerf360unboundedantialiased}, accessed from repositories on HuggingFace ~\cite{voxel51} and GitHub ~\cite{kerbl3Dgaussians}. Each scene was pretrained for 30,000 iterations using Inria's 3D Gaussian Splatting implementation and separately with Splatfacto to ensure robustness.

\begin{figure*}[t]
  \centering
  \includegraphics[width=1.0\textwidth]{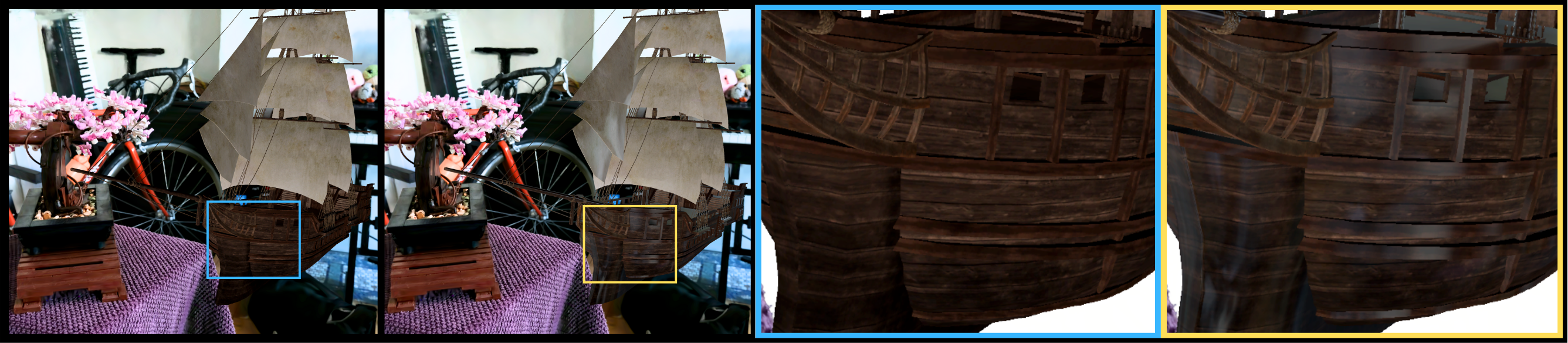}
  \caption{Comparison of a wooden ship model in a Gaussian-splatted scene. \textbf{Far Left:} Full scene without reflection probes. \textbf{Left Middle:} Full scene with GBake probes enabled. \textbf{Right Middle:} Close-up without reflection probes, showing flat appearance. \textbf{Far Right:} Close-up with probes, demonstrating improved environmental reflections and material detail.}
  \Description{A partially diffuse material wooden ship, the material is supposed to appear plastic but looks flat without the environmental lighting from the reflection probe.}
  \label{fig:ship}
\end{figure*}

For each scene, we compute bounding volumes, bake cubemaps, and relight mesh assets in Unity ~\cite{McGuire2017Data, unityasset}. Our results are most visually striking when applied to highly specular objects such as mirrors or chrome surfaces (Fig.~\ref{fig:teaser} and ~\ref{fig:room}). To demonstrate versatility, we also test materials with both diffuse and specular components. In Fig.~\ref{fig:ship}, we place a model ship asset adjacent to a Gaussian-splatted Lego bonsai tree, where the enabled GBake probes ensure consistent lighting conditions between the mesh and the splatted environment. All experiments were performed on an Intel Xeon Gold 6238R CPU at 2.20GHz with an NVIDIA 3090 GPU running CUDA 12.1. We include average runtimes for baking cubemaps of the Inria-trained scenes in a 5x5x5 grid at 800px resolution per face (Table ~\ref{tab:scene_comparison}).

\section{Conclusions and Future Work}
We create a method to bake 3D Gaussians for reflectance and environment mapping in Unity. We intend to integrate GBake with Unity's Universal Render Pipeline to bake cubemaps of both splats and meshes simultaneously. We would also like to revisit rasterization for faster seam-free baking and explore equirectangular projection for other spherical environment map representations.

\bibliographystyle{ACM-Reference-Format}
\bibliography{main}

\end{document}